\documentstyle[psfig]{l-aa}

\def\aeta{A\&A }
\def\aetal{A\&AL }
\def\aetas{A\&AS }
\def\apj{ApJ }
\def\apjs{ApJS }
\def\aj{AJ }
\def\mn{MNRAS }

\def\apjl{ApJL \rm}

\def\kms{km~s$^{-1}$\ }
\def\gsim{\lower.4ex\hbox{$\;\buildrel >\over{\scriptstyle\sim}\;$}}
\def\lsim{\lower.4ex\hbox{$\;\buildrel <\over{\scriptstyle\sim}\;$}}

\def\newpage{\vfill\eject}

\def\zabs{$z_{\rm abs}$}

\def\wr{$w_{\rm r}$}

\def\mgii{Mg~{\sc ii}}
\def\h{$h^{-1}_{100}$}

\begin{document}
\voffset -2.5truecm



\thesaurus{12.12.1;11.09.3;11.17.1}

\title{Physical conditions in metal line systems toward Q~1037-2704: 
Evidence for
superclustering at $z$~$\sim$~2\thanks{Based on observations
collected at the European Southern Observatory, La Silla, Chile.}
\thanks{Table~1 only available in electronic form
at the CDS via anonymous ftp to cdsarc.u-strasbg.fr (130.79.128.5)
or via http://cdsweb.u-strasbg.fr/Abstract.html}
 }
\author{Yannick Lespine\inst{1} 
\and Patrick Petitjean\inst{1,2}}
\offprints{P. Petitjean}
\institute{$^1$Institut d'Astrophysique de Paris - CNRS, 98bis Boulevard
Arago, F-75014 Paris, France\\
$^2$UA CNRS 173 - DAEC, Observatoire de Paris-Meudon, F-92195 Meudon
Principal Cedex, France}
\date{ }
\maketitle
\markboth{}{}
\begin{abstract}
We present a high spectral resolution optical spectrum
(FWHM~=~17~km~s$^{-1}$) of the bright quasar Q~1037-2704.
The large number of absorption systems suggests the presence
of an intervening supercluster, but the issue remains controversial.\par
We demonstrate that the strong C~{\sc iv} system at
$z_{\rm abs}$~$\sim$~2.08 spanning more than 1500~km~s$^{-1}$ is made up of
narrow components with typical Doppler parameter $b$~$\sim$~8~km~s$^{-1}$.
The gas is probably photo-ionized in which case the photo-ionizing 
spectrum should be dominated by stellar radiation. Analysis of
H~{\sc i}, Si~{\sc iv}, N~{\sc v} and C~{\sc iv} absorptions shows that the
gas has a fairly large metal content, $Z$~$>$~0.1~$Z_{\odot}$.\par
We detect C~{\sc ii}$^*$$\lambda$1335 absorption in subcomponents of
the $z_{\rm abs}$~$\sim$~2.133 system, which have moderate total 
C~{\sc ii} column
densities implying an electronic density of
$n_{\rm e}$~$\sim$~3~cm$^{-3}$. In addition, the weakness of the
O~{\sc i}$\lambda$1302, Si~{\sc ii}$\lambda$1304 and 
Fe~{\sc ii}$\lambda$1608
lines allows a detailed analysis of the line profiles. 
The results are that
the system has four narrow ($b$~$\sim$~10~km~s$^{-1}$) components; the
oxygen abundance is of the order of 0.03 solar; the iron abundance is
surprisingly large [Fe/O]~$\sim$~[Fe/O]$_{\odot}$~+~0.3. 
If photo-ionized by the
UV flux from the quasar, the distance between the absorber and the quasar
should be larger than 0.3$h^{-1}$~Mpc (with $h$ in units of 
$H_{\rm o}$~=~100~km~s$^{-1}$~Mpc$^{-1}$ and $q_{\rm o}$~=~0.5).\par
This demontrates that the gas is not associated with the quasar and, 
together with the detection of eight multiple metal systems 
between $z_{\rm abs}$~=~1.91 and 2.13,
suggests the presence of a coherent structure
extended over 80$h^{-1}$~Mpc along the line of sight. \par
\end{abstract}


\section{Introduction} \label{intr}
Evolution of large scale structures of the Universe is one of the most
important issues of modern cosmology.
QSO absorption line systems probe
material lying on the line of sight to quasars over a large
redshift range (0~$<$~$z$~$<$~5), and can thus be used as 
luminosity unbiased tracers of
these structures over most of the history of the Universe. \par
This has been recognized for over a decade (e.g. Shaver \& Robertson 1983,
Robertson \& Shaver 1983).
Observations of QSO pairs with projected separations from a few arcseconds to
a few arcminutes yield interesting constraints on the size,
physical structure and kinematics of
galactic haloes, clusters and filaments. Indeed, new
constraints have been obtained very recently on the extent
of the Ly$\alpha$ complexes perpendicular to the line of sight
at high (Smette et al. 1992, 1995, Bechtold et al. 1994, Dinshaw et al.
1994) and intermediate
(Dinshaw et al. 1995) redshifts indicating that they could
have sizes larger than 300~kpc. Such sizes are more indicative of
a correlation length than of real cloud sizes (Rauch \& Haenelt 1996).
This is consistent with the picture that the Ly$\alpha$ gas traces
the potential wells of dark matter filamentary structures (Cen et al.
1994, Petitjean et al. 1995, M\"ucket et al. 1996, Hernquist et al. 1996).
Large scale clustering of C~{\sc iv} systems (Heisler et al. 1989;
Foltz et al. 1993) or damped systems (Francis \& Hewett 1993, Wolfe 1993)
have also been detected recently.
The advent of 10m-class telescopes
will boost this field since faint QSOs in the same field
will allow 3-D mapping of the baryonic content of the universe via 
absorption line systems (Petitjean 1995).\par
It is well known that
formation of structures by gravitational instability depends
sensitively on the value of $\Omega$
and on the primordial density fluctuation spectrum. In particular,
observation of substructures in present day clusters (West 1994) may indicate
that the clusters have not finished evolving at $z$~$\sim$~0,
and favor high density universes ($\Omega > 0.5$, Richstone et al. 1992,
Jing et al. 1995) 
in contradiction with other determinations ($\Omega$~$\sim$~0.2-0.3, e.g.
Ostriker 1993). Therefore,
the search for high redshift clusters may yield important clues on
how structures form and subsequently merge, and so may be used to constrain
cosmological parameters (e.g. Evrard \& Charlot 1994). \par
In this context, the field surrounding the bright ($m_{\rm V}$~=~17.4)
high redshift ($z_{\rm em}$~=~2.193) QSO Tol Q1037-2704
is quite promising. Jakobsen et al. (1986) were the first to note
the remarkable similarity of the metal-line absorption systems in the
spectra of Tol~1037-2704 and Tol~1038-2712 separated by 17'9 on the sky,
corresponding to 4.3$h^{-1}_{100}$~Mpc for $q_{\rm o}$~=~0.5 at $z$~$\sim$~2.
They interpreted
this as evidence for the presence of a supercluster 
along the line of sight to the QSOs. The fact that the number of metal-line
systems in both spectra over the range 1.90~$\leq$~$z$~$\leq$~2.15
is far in excess of what is usually observed has
been considered as the strongest argument supporting this conclusion
(Ulrich \& Perryman 1986, Sargent \& Steidel 1987, Robertson 1987).
In a recent paper, Dinshaw \& Impey (1996, hereafter DI96) have presented 
new data on four 
quasars in this field. They find that the velocity correlation function
of the C~{\sc iv} systems shows strong and significant clustering 
for velocity separations less than 1000~km~s$^{-1}$ and 
up to 7000~km~s$^{-1}$ respectively. The spatial correlation function 
shows a marginally significant signal on scales of $<$~18~Mpc. 
They conclude that the dimensions of the proposed supercluster are at
least 30~$h^{-1}$~Mpc on the plane of the sky and approximately
80~$h^{-1}$~Mpc along the line of sight.\par
It is however not excluded that at least some of the systems,
in particular those spanning more than 1000~km~s$^{-1}$ at
$z_{\rm abs}$~$\sim$~2.08 and \zabs~$\sim$~2.13,
originate in gas ejected from the quasars. The fact that they are 
common to both Tol~1037-2704 and Tol~1038-2712 lines of sight
could be due to chance coincidence (Cristiani et al. 1987).\par
The importance of clarifying the presence of such a large structure
at $z$~$\sim$~2 led us to observe Tol~1037-2704
at a spectral resolution twice as high as previous observations, in
order to be able to discuss the physical state of the gas in
the \zabs~$\sim$~2.08 and \zabs~$\sim$~2.13 systems.
New optical data are presented in Section 2 and discussed in Section~3.
We draw our conclusions in Section~4.\par
\section{The data} \label{s2}
\subsection{Observations}
The observations were carried out at the F/8 Cassegrain focus of the 3.6~m
telescope at La Silla, ESO Chile. The spectra were obtained with the ESO
echelle spectrograph (CASPEC) during two observing runs:
three exposures of 5400~s each were obtained under good seeing
conditions in December 1992; two additional 5400~s exposures were taken
in April 1994 during cloudy nights.
A 300 line~mm$^{-1}$ cross disperser was used
in combination with a 31.6 line~mm$^{-1}$ echelle grating.
The detector was a Tektronix CCD with
568$\times$512 pixels of 27~$\mu$m square and a read--out noise of 10 electrons.
For each exposure
on the object flat field images and wavelength comparison Thorium--Argon
spectra were recorded. The slit width was 2" corresponding to a
spectral resolution of $R$~$\sim$~18000.
The accuracy in the wavelength
calibration measured on the calibrated Thorium--Argon spectra is
about 0.03~\AA. \par
The data were reduced using
the echelle reduction packages provided by IRAF and MIDAS.
The cosmic--ray events have been removed in the regions between object
spectra before extraction of the object.
Several exposures were co--added to increase the signal to noise
ratio. During this merging procedure the cosmic--ray events affecting
the object pixels were recognized and eliminated, a  procedure that worked well
since five exposures were available.
The background sky spectrum
was difficult to extract separately due to the small spacing between
the orders in the blue.
Instead, we have carefully fitted the zero level to the bottom of the
numerous saturated
lines in the Ly$\alpha$ forest. The uncertainty on the determination
can be estimated to be 5\%. The normalized spectrum is shown in Fig.~1. \par
We identify all the absorption features with equivalent widths larger 
than 5$\times$FWHM$\times$$\sigma$ where $\sigma$ is the noise rms in the 
adjacent continuum and use the linelist to recognize the systems.
A system 
is identified when the wavelength agreement beetween two lines of a
doublet found beyond the Ly$\alpha$ forest is better than 0.1~\AA~ and
when the strengths of the two lines are consistent with the 
corresponding oscillator strength ratio. All the systems listed in 
Section~2.2 have doublets redshifted beyond the Ly$\alpha$ forest.
The whole spectrum is then searched for additional lines. For isolated
lines, the same criterion on the wavelength is used. For strong blends,
the identification is based on the consistency of the line profiles 
(see e.g. Fig.~2). 
All features detected in the spectrum are described
in Table 1 and identification is given when possible. 
We then have fitted Voigt profiles convolved with the instrumental profile to
all the identified absorption lines using the package FIT/LYMAN recently
implemented in MIDAS. The atomic data
are taken from Morton et al. (1988), Morton (1991) and Tripp et al. (1996).
The systems are considered made up of components. For each of the latter 
the redshift, temperature, and 
turbulent Doppler parameter are assumed the same for the high-ionization
(C~{\sc iv}, Si~{\sc iv}) and low-ionization (C~{\sc ii}, O~{\sc i}, 
Al~{\sc ii}, Si~{\sc ii}, Fe~{\sc ii}) species.
The model is fitted to all the absorption lines of the system at the same 
time minimizing the $\chi^2$ and the number of subcomponents.
In the case large Doppler parameters are found, especially in the centre of
strong blends, it is probable that
several subcomponents are present. As shown by Jenkins (1986), 
the Doppler parameter is then indicative of the number of subcomponents and,
if the lines are not strongly saturated, the column density 
is a good estimate of the sum of the column densities in all subcomponents.
The case of the $z_{\rm abs}$~=~2.081 system is peculiar as 
tight constraints on the Doppler parameter can be placed because some of 
the lines are saturated but do not go to the
zero level. In this system we find $b$~$\la$~8~km~s$^{-1}$. 
The vacuum heliocentric redshifts,
velocity dispersions and column densities of subcomponents
are given in Table 1. The typical error on the column densities is 0.15 in 
units of log~$N$ except for heavily blended lines marked in the Table.
\par\medskip\noindent
\begin{figure*}
\centerline{\vbox{
\hbox{\psfig{figure=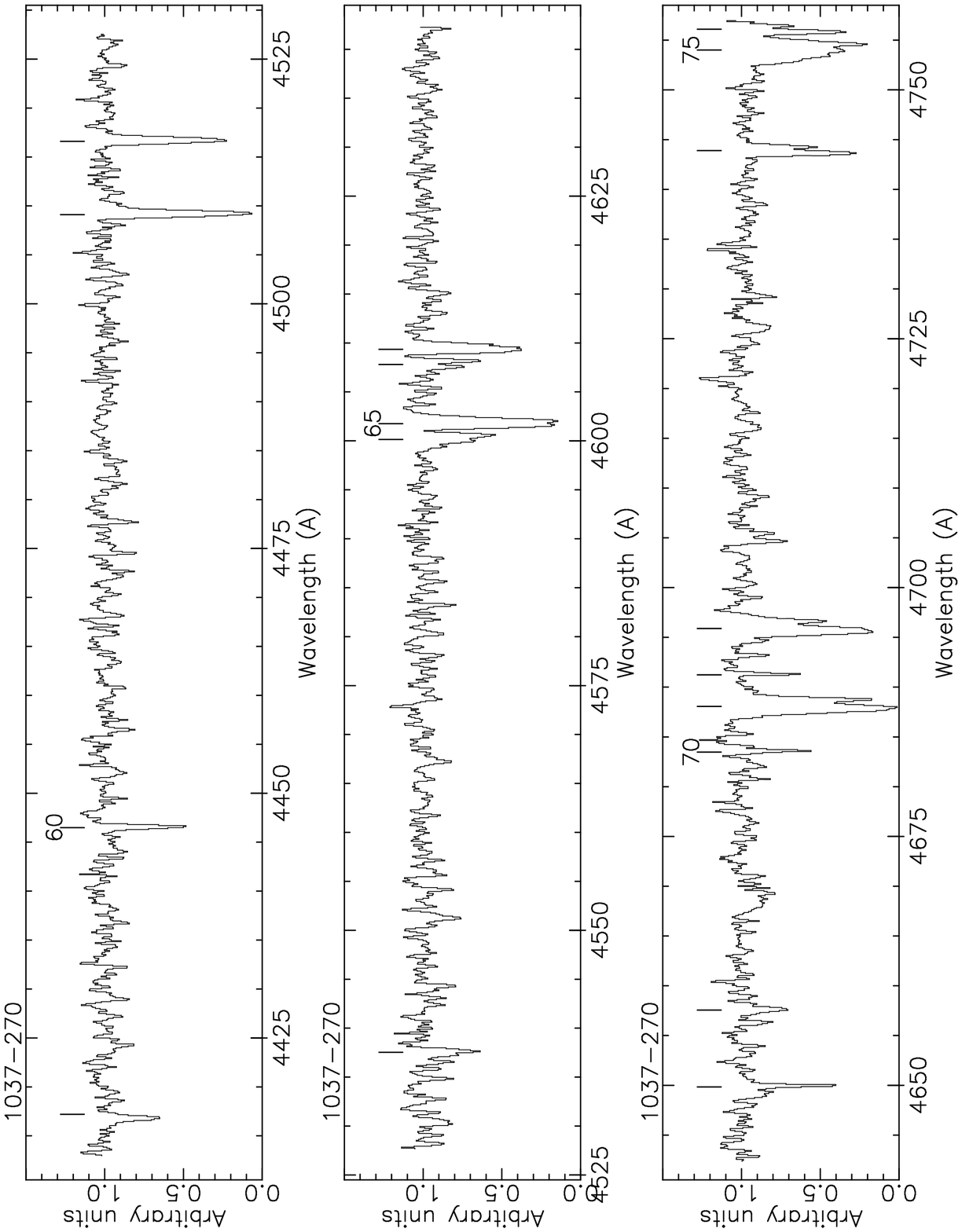,height=11.cm,angle=0}
\psfig{figure=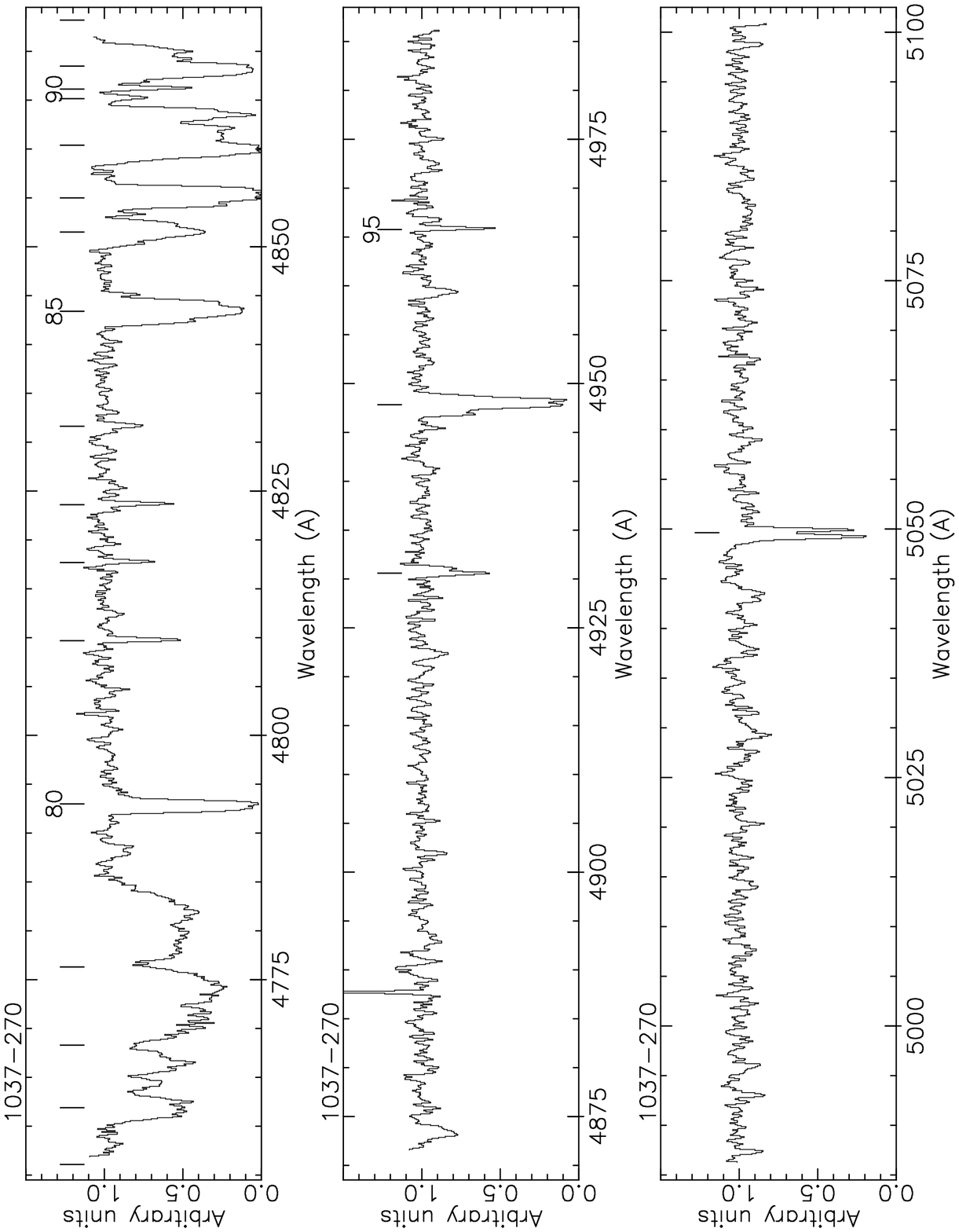,height=11.cm,angle=0}}
\hbox{\psfig{figure=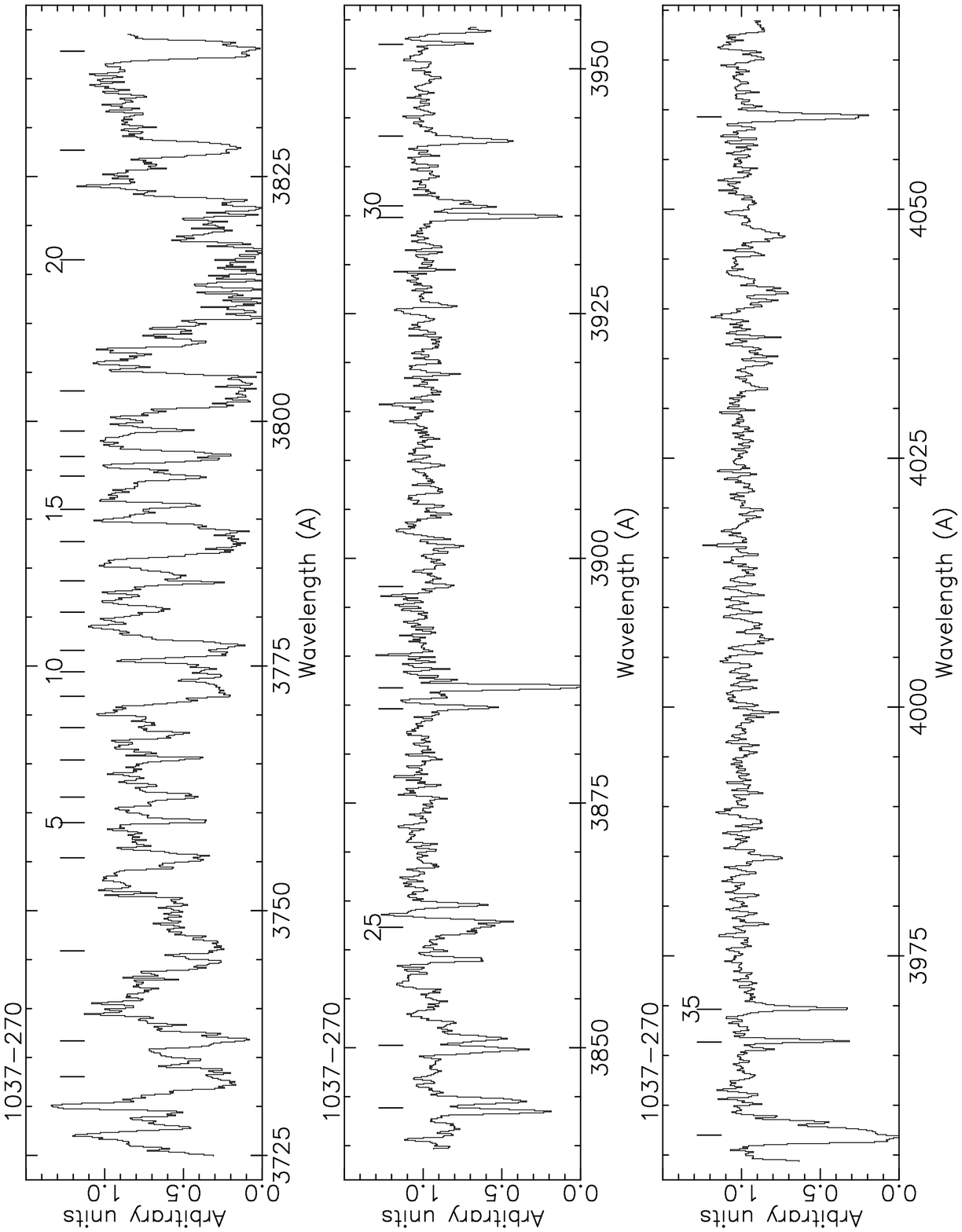,height=11.cm,angle=0}
\psfig{figure=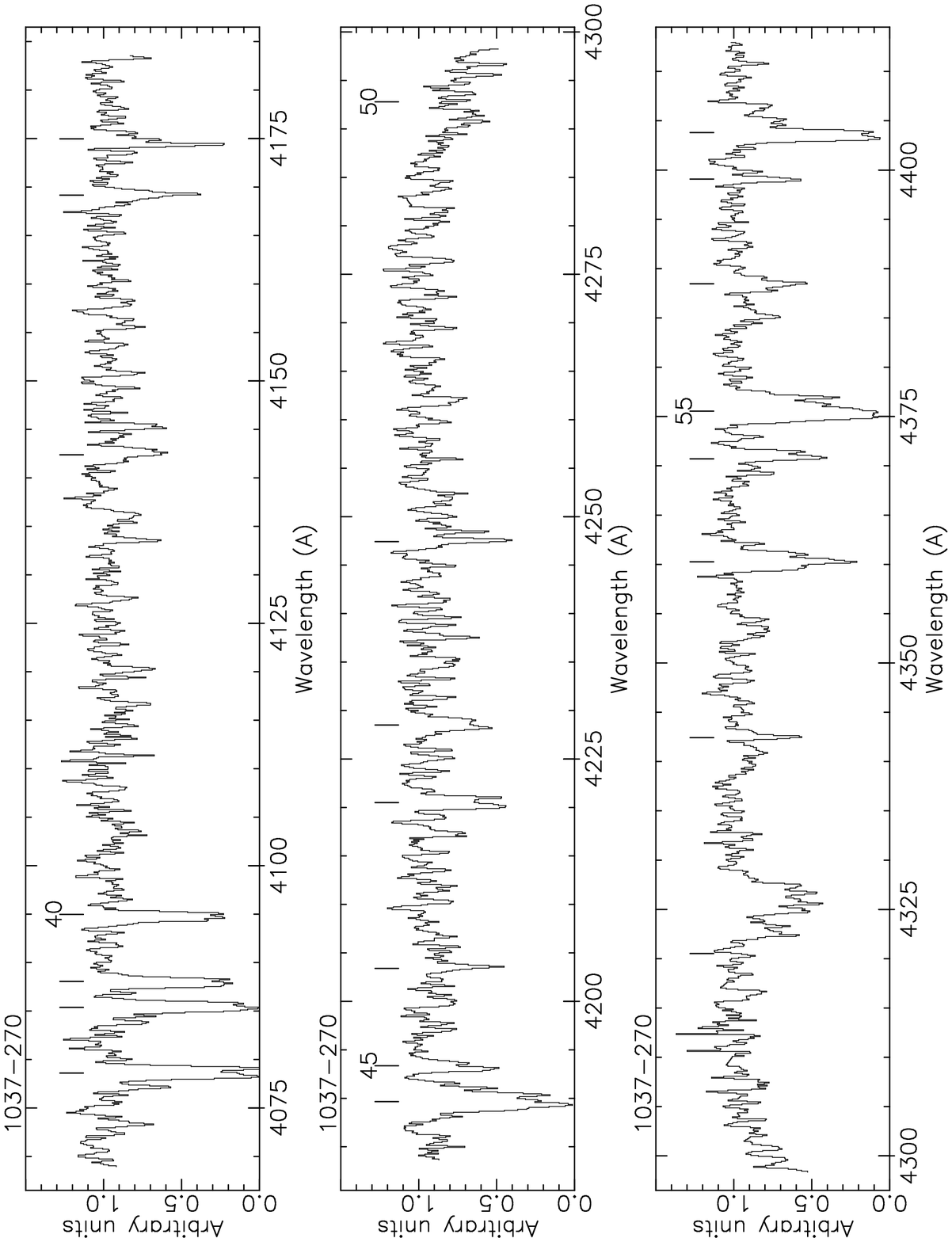,height=11.cm,angle=0}}
}}

\caption[]{Normalized spectrum of Tol~1037-270}
\end{figure*}
%
%
%
\subsection{Comments on individual systems} \label{s3}
\subsubsection{Galactic absorption}
We detect a galactic Ca~{\sc ii}$\lambda\lambda$3934,3969 doublet
with equivalent widths 0.48, 0.30~\AA~ and
heliocentric velocity $v$~=~3.3~km~s$^{-1}$. The column density
and Doppler parameter are 10$^{13.2}$~cm$^{-2}$
and 12.6~km~s$^{-1}$.
\subsubsection{z$_{\rm abs}$~=~0.6964} \label{s31}
This \mgii~ system, detected at the 5$\sigma$ level, shows two resolved
components separated by about 35~\kms.
Fe~{\sc ii}$\lambda$2382 is not detected and should have
$w_{\rm r}$~$<$~0.11~\AA. It is interesting to note that
Jakobsen \& Perryman (1992) detect a strong Mg~{\sc ii} system at
\zabs~=~0.643 in BW15 (see Bohuski \& Weedman 1979).
The 3D distance between the two systems is about 200\h~Mpc at $z$~=~0.696.
\subsubsection{z$_{\rm abs}$~=~1.0765} \label{s32}
We detect Fe~{\sc ii}$\lambda\lambda\lambda$2344,,2374,2382 lines in this
Mg~{\sc ii} system (Jakobsen et al. 1986).
At our resolution, the line profiles indicate the presence of
three components spread over 72~\kms.
\subsubsection{z$_{\rm abs}$~=~1.4829} \label{s33}
The redshift coincidence of the two C~{\sc iv} lines in this system
is very good. There are
two detached components with velocity separation $\Delta V$~$\sim$~97~\kms.
The corresponding Al~{\sc iii} lines have \wr~$<$~0.04~\AA.
\subsubsection{z$_{\rm abs}$~=~1.6345} \label{s34}
This system has two strong components separated by 50~\kms
and an additional detached satellite displaced 100~\kms to the
blue. The corresponding Al~{\sc iii} lines have \wr~$<$~0.05~\AA.
Considering the lines as optically thin, this corresponds to
$N$(Al~{\sc iii})~$<$~3$\times$10$^{12}$~cm$^{-2}$. Given the
C~{\sc iv} column density derived from line profile fitting
(see Table~1), we obtain $N$(C~{\sc iv})/$N$(Al~{\sc iii})~$>$~86.
This is four times larger than the maximum of this ratio observed in the
Galactic halo (Sembach \& Savage 1992).
\subsubsection{z$_{\rm abs}$~=~1.9125} \label{s35}
This is a strong system with very good wavelength coincidences. 
Si~{\sc iv}$\lambda$1402 is blended with C~{\sc iv}$\lambda$1550 at
\zabs~=~1.6345. We detect strong absorptions by C~{\sc ii}$\lambda$1334,
Si~{\sc ii}$\lambda$1526 and Al~{\sc ii}$\lambda$1670. 
The fit with one component is very good. However the strength of 
C~{\sc ii}$\lambda$1334 suggests the presence of subcomponents.\par
Using Bergeron \& Stasi\'nska (1986) and Petitjean et al. (1994)
models, we can estimate the characteristics of the
gas. 
From log~$N$(Si~{\sc ii})/$N$(Si~{\sc iv})~$\sim$~0.1 and
log~$N$(C~{\sc ii})/$N$(C~{\sc iv})~$\sim$~0.6,  we derive
an ionization parameter $U$ close to 2$\times$10$^{-3}$.
Using \wr~$\sim$~0.9~\AA~ for the H~{\sc i} Ly$\alpha$ line
(Sargent \& Steidel 1987) we find 10$^{17}$~cm$^{-2}$
for $b$~$<$~40~km~s$^{-1}$. From this 
we can conclude that all the column densities are indicating
abundances close to but smaller than 0.1~$Z_{\odot}$. 
Note that the Al~{\sc ii} column density is consistent with this conclusion.
This is an indication 
that the system is not associated with the quasar (Petitjean et al. 1994).
\subsubsection{z$_{\rm abs}$~=~1.9722} \label{s36}
This strong C~{\sc iv} system is a blend of four components spread over
130~\kms; three of them have an associated Si~{\sc iv} absorption.
The strongest component has narrow ($b$~$<$~10~\kms) 
and well defined
C~{\sc ii}, Si~{\sc ii} and Al~{\sc ii} absorption lines.
The column density ratios (see Table~1) are similar to that 
obtained for the \zabs~=~1.9125 system. The H~{\sc i} column density
should be larger than 10$^{18}$~cm$^{-2}$, from the Ly$\alpha$ line
equivalent width (Sargent \& Steidel 1987). 
The abundances are therefore most certainly smaller than 0.1~$Z_{\odot}$
as in the z$_{\rm abs}$~=~1.9125 system.
\subsubsection{z$_{\rm abs}$~=~2.0034} \label{s37}
This is a weak (\wr(1548,1550)~=~0.27, 0.15~\AA) C~{\sc iv} system.
No other line is detected.
\subsubsection{z$_{\rm abs}$~=~2.0280} \label{s38}
Two strong components separated by 50~\kms~ and a weak satellite 
275~\kms~ to the blue are seen in this system. The weak detached lines
have $b$~$<$~10~\kms~ consistent with photo-ionization. 
C~{\sc ii}$\lambda$1334 is at our
detection limit (see DI96). We do not detect the component at 
$z_{\rm abs}$~=~2.03915 (DI96).
\subsubsection{z$_{\rm abs}$~$\sim$~2.081} \label{s39}
This redshift is common to both Tol~1037-2704 and Tol~1038-2712 lines 
of sight (DI96). The system shows absorption spread over the 
redshift range 2.0699~$<$~$z$~$<$~2.0879 or 1750~km~s$^{-1}$. 
The C~{\sc iv} lines are weakly saturated (the equivalent widths ratio is 
different from the oscillator strength ratio) but do not go to the
zero level (see Section \ref{s41}). The Ly$\alpha$ and Si~{\sc iv} lines (at 
3748 and 4294~\AA~ respectively) are weak. 
We cannot exclude
weak N~{\sc v} absorption since strong H~{\sc i} and Si~{\sc iii} 
lines at $z_{\rm abs}$~=~2.133 are present in the redshifted N~{\sc v} 
wavelength range.  
Even at our resolution the C~{\sc iv} profile does not break into components 
suggesting absorption by a continuous medium. 
The appearance of such complex
may lead Tol~1037-2704 to be classified as a BAL QSO (Weymann et al.
1981). This is the reason why a detailed discussion of the physical 
state of the absorbing gas is useful (see Section 3.1).
\subsubsection{z$_{\rm abs}$~=~2.1067 and 2.1157} \label{s310}
These are weak (\wr(1548,1550)~=~0.07, 0.04~\AA~ and 
0.06, 0.04~\AA) C~{\sc iv} systems with a single narrow 
($b$~$\sim$~10~km~s$^{-1}$)
component. No other line is detected.
\subsubsection{z$_{\rm abs}$~$\sim$~2.133} \label{s311}
The C~{\sc iv} absorption is spread over the redshift range 
2.1279~$<$~$z$~$<$~2.1403 or 1190~km~s$^{-1}$. It shows well detached 
saturated complexes. 
The line profile is complex but not as smooth as
the $z_{\rm abs}$~=~2.081 profile. Visually there are two well
defined components 
at $z_{\rm abs}$~=~2.1392 and 2.1396 seen in O~{\sc i}$\lambda$1302, 
Si~{\sc ii}$\lambda$1304, Fe~{\sc ii}$\lambda$1608 and 
C~{\sc ii}$^*\lambda$1335 (see Section 3.2).
DI96 identified the $\lambda$4193.4 feature as C~{\sc ii}$\lambda$1334
at $z_{\rm abs}$~=~2.14233 on the basis of the presence of 
C~{\sc iv}$\lambda\lambda$1548,1550 and Si~{\sc iv}$\lambda$1393 at the same
redshift. However, in the DI96 data, some of these features are blended with
other lines. At our higher spectral resolution,
we find that, if the C~{\sc ii}$\lambda$1334
identification is correct, the 
redshift disagreement with C~{\sc iv} is about 30~km~s$^{-1}$ that is
larger than our resolution.
Moreover, at this redshift, we see neither Si~{\sc iv}$\lambda$1393 nor 
Si~{\sc ii}$\lambda$1260 
which is as strong as C~{\sc ii}$\lambda$1334 in neighbouring 
components. 
On the contrary, the identification as C~{\sc ii}$^*\lambda$1335
gives nearly perfect redshift agreement with O~{\sc i}, Si~{\sc ii} and
Fe~{\sc ii} (see Fig.~2). We therefore
believe that the identification as C~{\sc ii}$^*\lambda$1335
is secure. This makes the detailed studying of this system of great importance
for discussing the origin of the gas (see Section~3.2).\par
It is interesting to note that a narrow and single 
N~{\sc v}$\lambda$1238 line is detected at $z_{\rm abs}$~=~2.1358.
N~{\sc v}$\lambda$1242 is at our detection limit.
\begin{figure}
\centerline{\vbox{
\psfig{figure=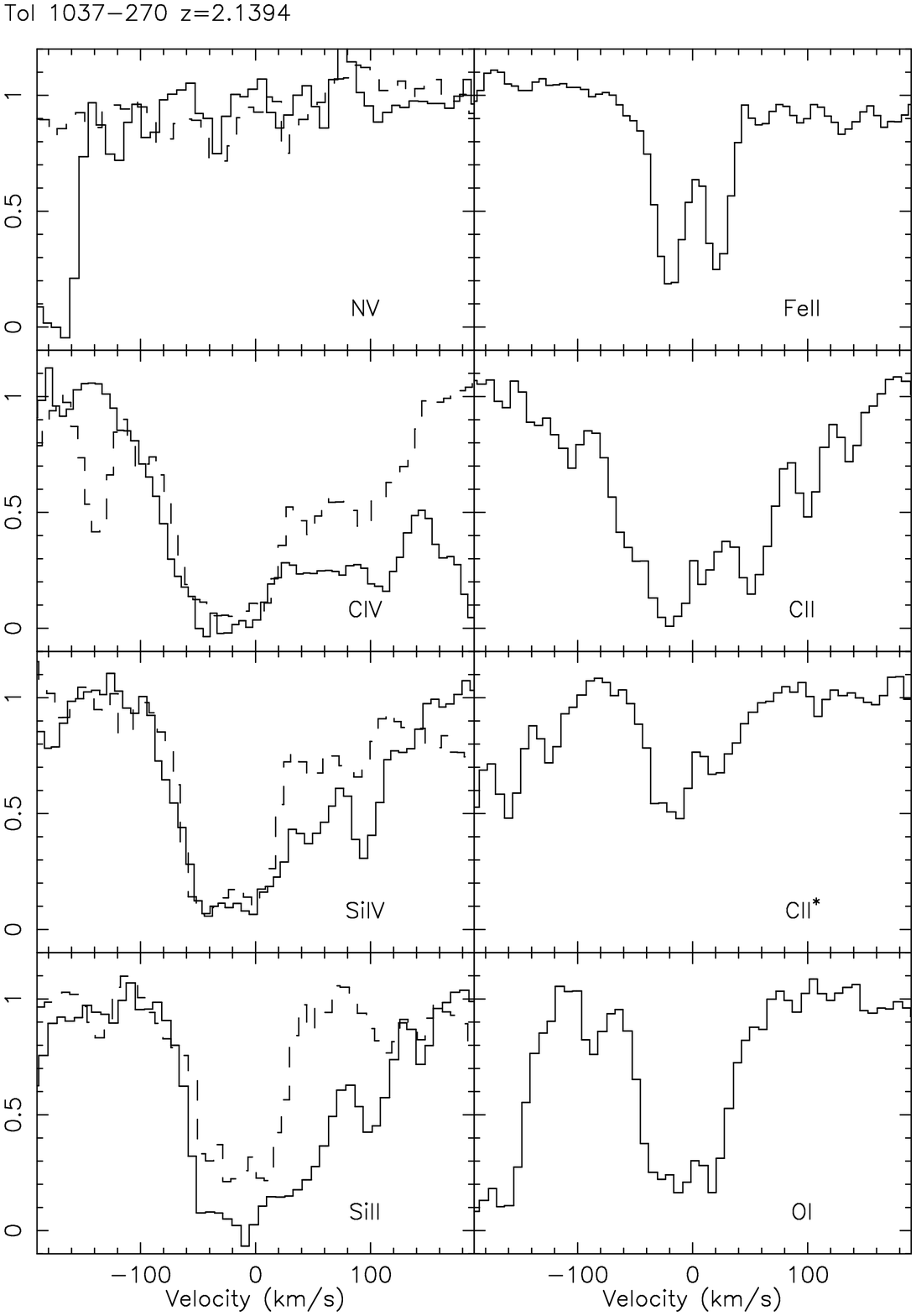,height=12.cm}
}}
\caption[]{Absorption lines in the \zabs~$\sim$~2.133 system
on a relative velocity scale. The redshift for zero velocity is given
at the top. When several lines of the same ion are present in the data
the corresponding parts of the spectrum are overplotted, with solid,
dashed and dotted lines in order of decreasing oscillator strengths.
CII is for CII$\lambda$1334, CII$^*$ for CII$\lambda$1335, CIV for
CIV$\lambda$$\lambda$1548,1550, OI for OI$\lambda$1302,
SiII for SiII$\lambda$$\lambda$1260,1304, 
SiIV for SiIV$\lambda$$\lambda$1393,1402, FeII for FeII$\lambda$1608
}
\end{figure}
%
%
%
%
%
\psscalefirst
\begin{figure*}
\picplace{22.5 cm}
\centerline{\vbox{
\hbox{
\psfig{figure=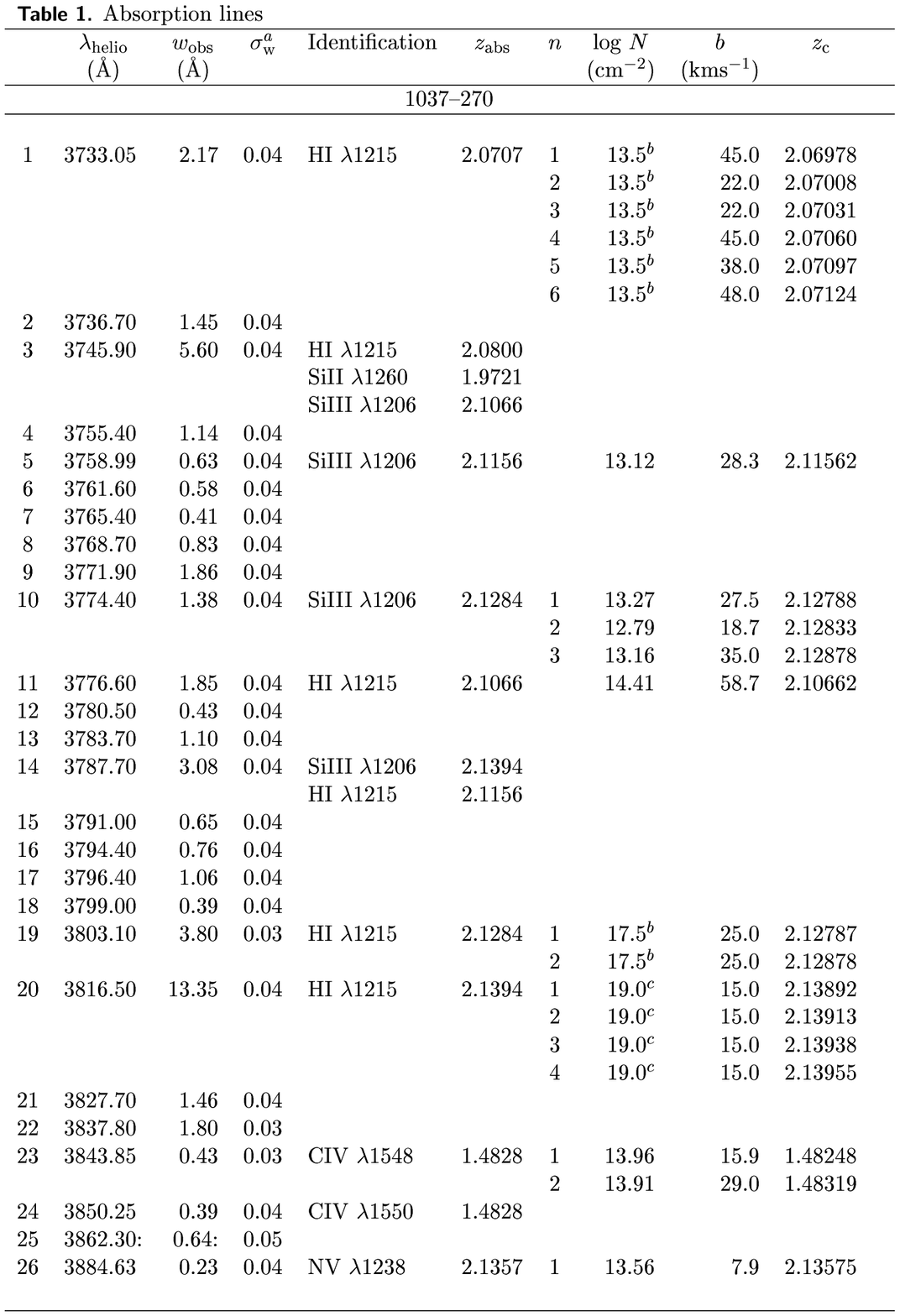,height=11.cm,angle=0}
\psfig{figure=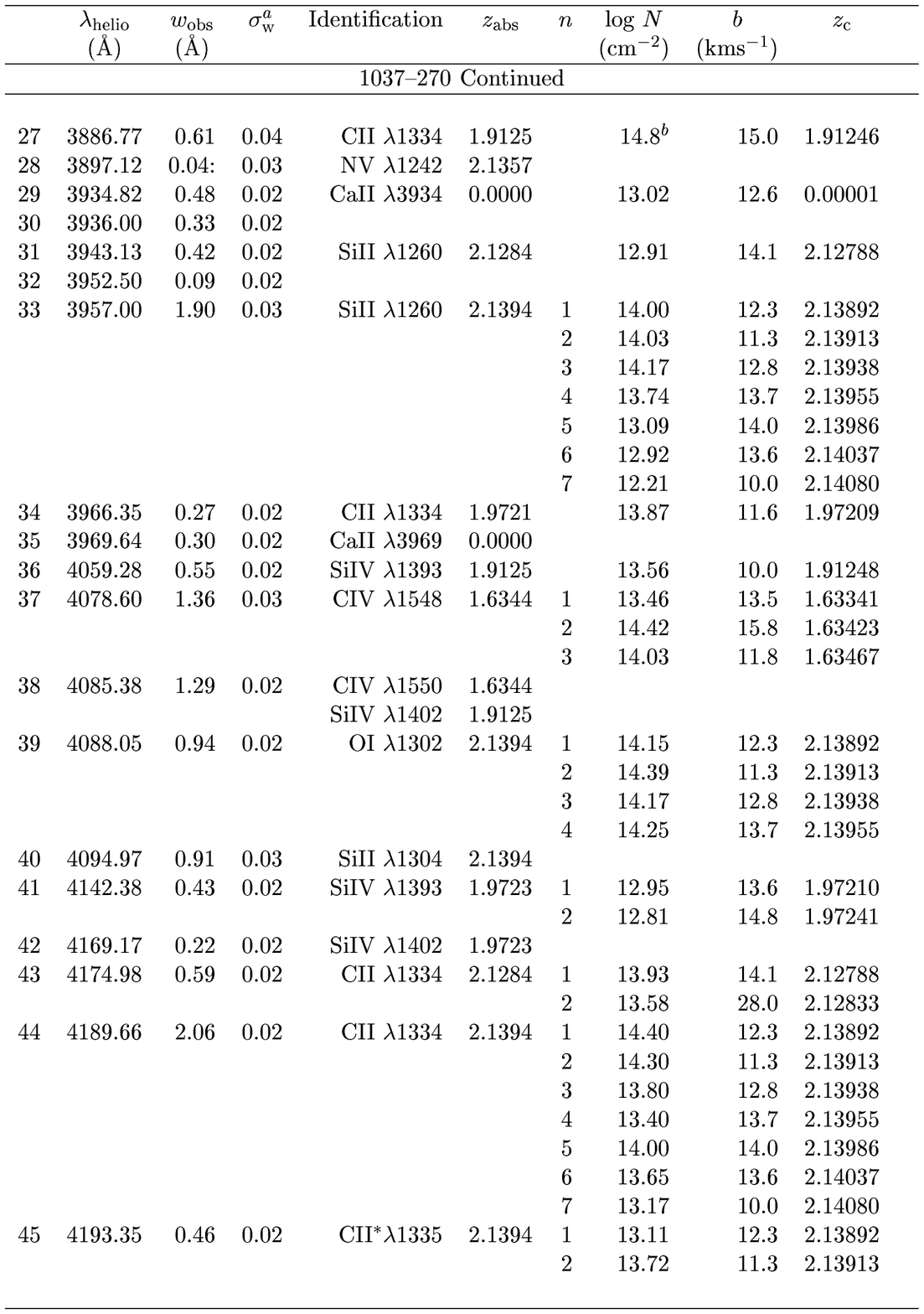,height=11.cm,angle=0}}
\hbox{
\psfig{figure=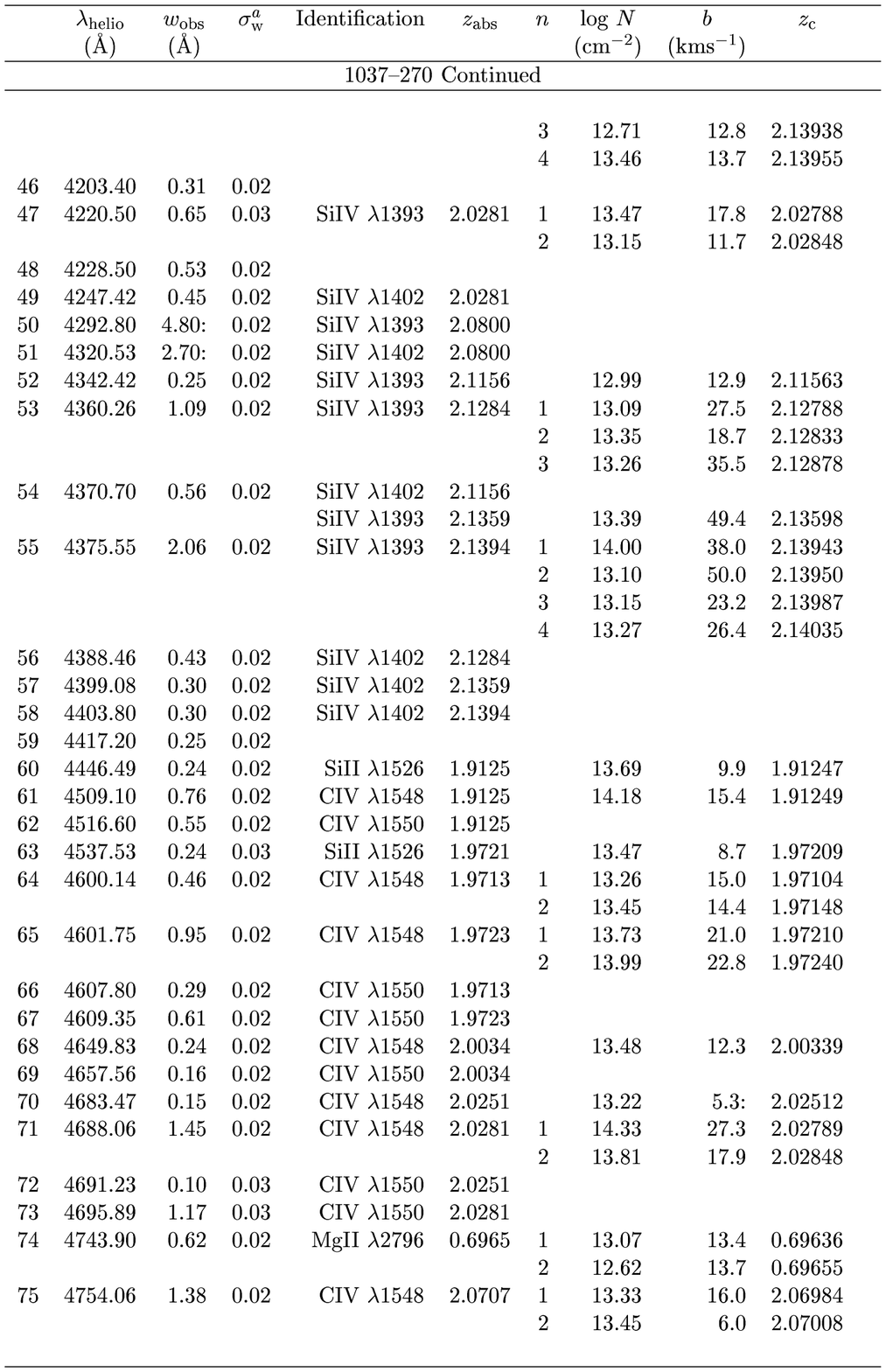,height=11.cm,angle=0}
\psfig{figure=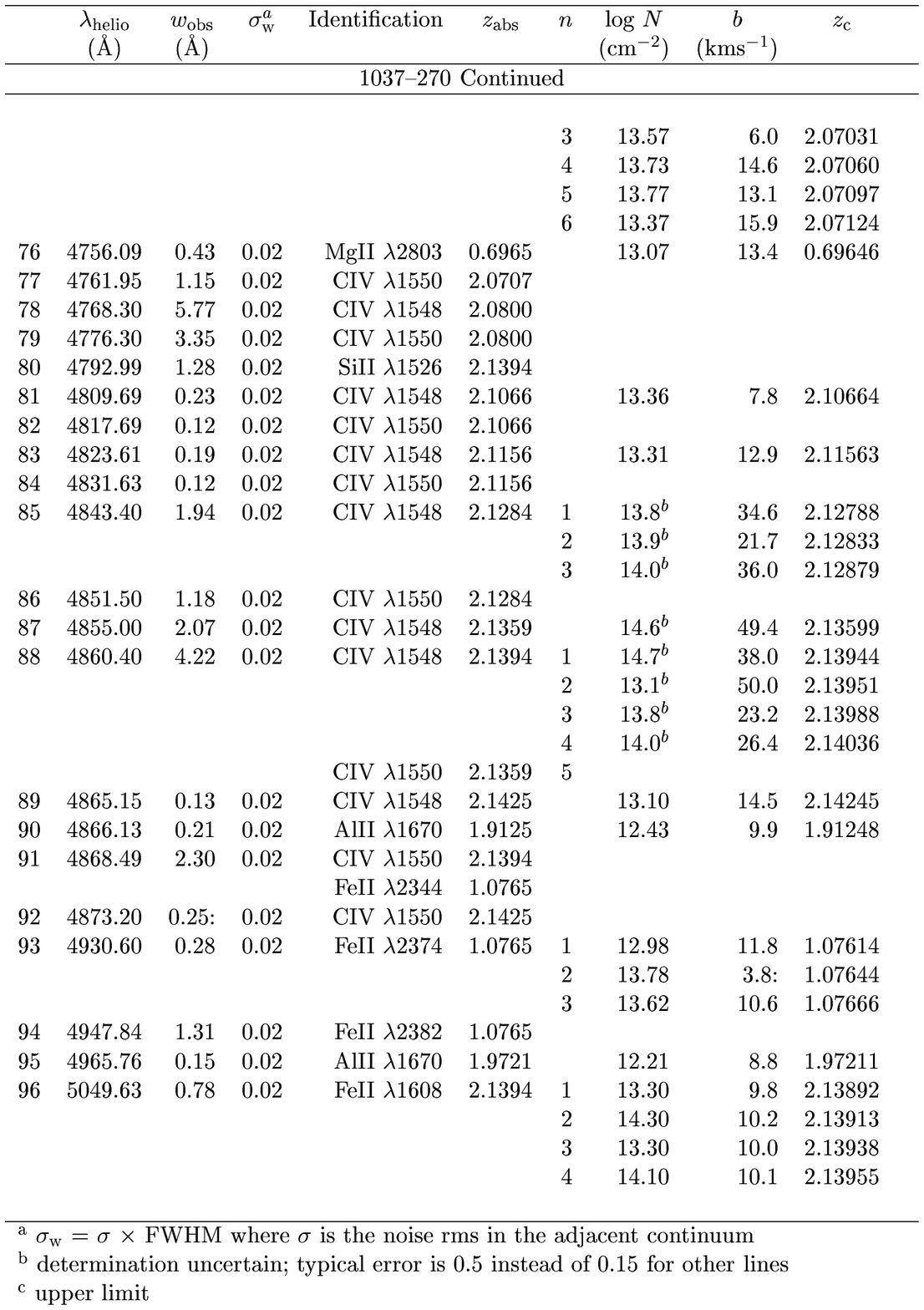,height=11.cm,angle=0}}
}}
\caption[]{}
\end{figure*}
\psrotatefirst

\section{Physical state of the gas} \label{s4}
To ascertain the possible presence of a large concentration of gas 
possibly extended
over 30$h^{-1}$~Mpc perpendicular to the line of sight and 80$h^{-1}$~Mpc 
along the line of sight (DI96), we use our data to make
a detailed analysis of the $z_{\rm abs}$~=~2.081
and 2.133 systems to differentiate between the ejection
and the intervening hypotheses for the origin of these systems.
\begin{figure}
\centerline{\vbox{
\psfig{figure=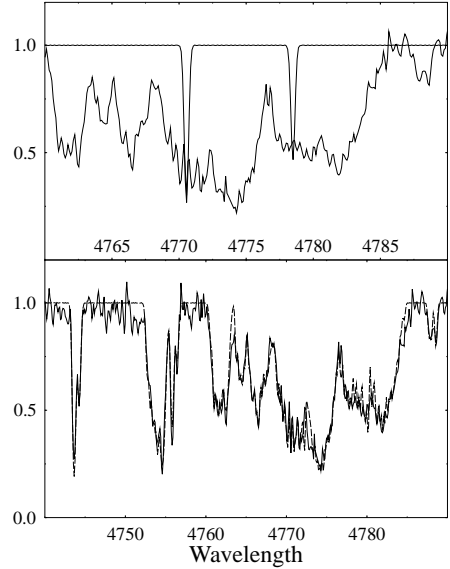,height=8.cm}
}}
\caption[]{Fit to the C~{\sc iv} absorption at \zabs~=~2.081.
In the upper panel a one component C~{\sc iv} doublet with 
$b$~=~7~km~s$^{-1}$ and log~$N$~=~13.65 is overplotted on the spectrum.
This illustrates the necessity to use narrow components to fit the
data. 
The overall fit, where subcomponents with
4~$<$~$b$~$<$~10~km~s$^{-1}$ are used, is shown in the lower panel.
}
\end{figure}
\subsection{The $z_{\rm abs}$~$\sim$~2.081 system} \label{s41}
It can be seen in Fig.~3 that the C~{\sc iv} lines do not go to the
zero level. Most of the components have apparent optical depth smaller 
than 0.7. It is therefore expected that the apparent optical depth ratio 
$\tau(\lambda1548)$/$\tau(\lambda1550)$
is approximately equal to the 
oscillator strength ratio, $f$($\lambda$1548)/$f$($\lambda$1550)~=~2, 
if the lines
are resolved. It is apparent that this is not true:
$\tau(\lambda1548)$/$\tau(\lambda1550)$~$<$~1.5. This suggests either that
the absorbing cloud do not cover the continuum or that the cloud
is made up of unresolved components.
Several arguments favor the later assumption.
The velocity difference between the absorber and the quasar 
($z_{\rm em}$~=~2.193, Sargent \& Steidel 1987) is about
10000~km~s$^{-1}$, with no trace of similar absorption at smaller 
velocity difference 
(see Section 3.2 for discussion of the $z_{\rm abs}$~$\sim$~2.133 system).
In order to partially cover the continuum source, the clouds
must be small and close to the quasar. 
However, the system is common to Tol~1037-2704 and Tol~1038-2712
(DI96) which are separated by 4$h^{-1}$~Mpc, and it is very unlikely that 
a sheet of very small clouds would extend from one QSO to the other or that
by chance material ejected by two independent QSOs would have the same 
velocity difference (Sargent \& Steidel 1987). 
Finally, the system is not very highly 
ionized contrary to what is expected for gas close to the quasar 
(see typical associated systems in Petitjean et al. 1994).\par
Assuming that the system is made up of narrow components, 
we have fitted the C~{\sc iv} line profile. Since we do not
resolve individual components, the fit cannot be unique. We assume
therefore that individual components have Doppler parameters
in a given range $b_{\rm i}$--$b_{\rm s}$. We evaluate the mean 
Doppler parameter by
fitting the optical depths of both lines of the doublet
(see Fig.~3 top panel) and find $b$~$\sim$~8~km~s$^{-1}$.
We have checked that the best fit is obtained with $b_{\rm i}$~=~4
and $b_{\rm s}$~=~10~km~s$^{-1}$.
The global fit is given in Fig.~3 (lower panel). 
We conclude from this that the gas temperature is smaller than 10$^5$~K 
and probably smaller than 5$\times$10$^{4}$~K. 
Using the positions and Doppler parameters of the components derived 
from the fit, we then have 
fitted the Si~{\sc iv}, H~{\sc i} and N~{\sc v} wavelength ranges 
to derive column densities or upper limits.\par
The conditions are fairly homogeneous in the complex
and typical column densities are log~$N$~$\sim$~13.2, 12.3, 13., 13.7 and 
$<$~13 for H~{\sc i}, Si~{\sc iii}, Si~{\sc iv}, C~{\sc iv}
and N~{\sc v} respectively. The column density ratios are incompatible
with ionization by a QSO-like spectrum. Indeed,
from the large $N$(Si~{\sc iv})/$N$(C~{\sc iv})
ratio, the ionization parameter should be $U$~$\sim$~10$^{-3}$ (see 
Bergeron \& Stasi\'nska 1986) whereas the large 
$N$(Si~{\sc iv})/$N$(Si~{\sc iii}) ratio indicates $U$~$\sim$~0.1.
One way to solve the contradiction is to consider that the gas is
photo-ionized by a stellar spectrum with no photons above 45~eV.
This suggests that this system originates in a H~{\sc ii} region 
(see also Viegas \& Gruenwald 1991).
It should be then highly ionized with 
Si~{\sc iv}/Si~$\sim$~C~{\sc iv}/C~$\sim$~1. In this case 
since 10$^5$ is a reasonable upper limit for H~{\sc ii}/H~{\sc i}, the 
abundances should be larger than 0.1~$Z_{\odot}$.
Alternatively if the temperature is $T$~$\sim$~10$^5$~K and the gas
collisionally ionized, the ionization correction 
(see Arnaud \& Rothenflug 1985)
leads to carbon and silicon abundances of the order of 0.5 and 5 times
the solar value respectively. 
Although we cannot firmly differentiate between both explanations, 
we can conclude that the gas is not photo-ionized by the QSO and has
fairly high metallicity for such redshift.  
\subsection{The $z_{\rm abs}$~$\sim$~2.133 system} \label{s42}
Although the C~{\sc iv} lines are particularly strong in this system, the
line profiles are typical of a normal intervening system. In particular 
there is a strong H~{\sc i} absorption and the
low-ionization species are spread over a velocity range much smaller
than C~{\sc iv}. $\Delta V$~=~40, 40, 1200, 1200, 2200 and 3000~\kms~  
for O~{\sc i}, Fe~{\sc ii}, Si~{\sc ii}, C~{\sc ii}, Si~{\sc iv} and 
C~{\sc iv} respectively. \par
The C~{\sc iv} profile is complex and heavily blended, so we concentrate 
our analysis on the O~{\sc i}, C~{\sc ii}$^*$, C~{\sc ii},
Si~{\sc ii} and Fe~{\sc ii} lines (see Fig.~2).
The fit of the lines has been performed assuming that the 
redshift and $b$ values were the same for all species which is reasonable
since the absorptions are expected to arise from the same phase.
Since we see two well detached components in Fe~{\sc ii}, we first 
tried to find a two-component model. This failed because the $b$ values
mostly constrained by Fe~{\sc ii} were too large ($\sim$20~km~s$^{-1}$) for
the Si~{\sc ii}$\lambda$1304 fit to be acceptable. The obvious solution
was that the system is made up of a larger number of narrower components.
We find a good overall fit for O~{\sc i}, C~{\sc ii}$^*$, C~{\sc ii} and
Si~{\sc ii} with four $b$~$\sim$~12~km~s$^{-1}$ components. The
Fe~{\sc ii} profile imposes $b$~$\sim$~10~km~s$^{-1}$. This can be understood
if the lines arise in a medium with turbulent motions characterised
by $b$~$\sim$~8~km~s$^{-1}$ and $T$~$\sim$~50000~K. 
This values should be considered as upper limits. \par 
We find that the total column densities are 
log~$N$(O~{\sc i}, C~{\sc ii}$^*$, C~{\sc ii},
Si~{\sc ii} and Fe~{\sc ii})~=~14.9, 14.0, 14.8, 14.6, 14.6 respectively.
If the C~{\sc ii} column density can be considered as
quite uncertain since it relies on a single strong line, this is not the 
case for other species which have at least one weak line.
The H~{\sc i} Ly$\alpha$ line is redshifted in a part of the spectrum
with moderate S/N ratio. However from the lack of damped wings (see
Fig.~1 and DI96), we are confident that the total H~{\sc i} column
density is smaller than 4$\times$10$^{19}$~cm$^{-2}$. \par
We discuss the metallicities in the system compared to the logarithmic
solar values: --3.4, --3.2, --4.6 and
--4.5 for C, O, Fe and Si respectively. O~{\sc i} and H~{\sc i} being
tied up by charge-exchange reaction, the $N$(O~{\sc i})/$N$(H~{\sc i})
column density ratio gives the oxygen abundance without need of any 
ionization correction. We thus conclude that the oxygen abundance is 
of the order of 0.03 solar.\par
Scaling photo-ionization models published in the literature 
(see Steidel 1990, Petitjean et al. 1994), we can derive relative abundances
for other species. The above column densities indicate that the ionization
parameter is smaller than 10$^{-3}$. For log~$N$(H~{\sc i})~=~19.5,
and metallicities $Z$~=~0.01~$Z_{\odot}$, the models give log~$N$(O~{\sc i}, 
C~{\sc ii}, Si~{\sc ii} and Fe~{\sc ii})~$\sim$~14.5, 15.1, 14.1 and 13.6 
respectively. 
It can be seen that iron could be  
overabundant compared to oxygen by a factor of two to three. \par
The conclusion that the [Fe/O] 
ratio is close to solar for an absolute abundance of the order of
a hundredth of solar is intriguing. 
Indeed observations of metal deficient stars in the halo of our
galaxy indicate that for [Fe/H]~$<$~--2, [O/Fe]~$>$~0.5 (Spite 1991),
consistent with predictions of chemical evolution models (Matteucci
\& Padovani 1993). \par
The detection of the C~{\sc ii}$^*$$\lambda$1335 line shows that the
fine structure upper level of the C~{\sc ii} ground-state is populated.
If this is due to collisions and if it is assumed that the gas is
completely ionized as suggested by the moderate H~{\sc i} column density
(H~{\sc i}/H~$<$~1), 
one can derive a value $n_{\rm e}$~$\sim$~3~cm$^{-3}$ for the electronic 
density. 
We can combine the density and the ionization parameter
derived from the ionization state of the gas to infer information about the 
ionizing flux.
Using $F_{\nu}$~=~$F_{\rm o}$($\nu$/$\nu_{\rm o}$)$^{-\alpha}$ 
with $\alpha$~=~0.5 or 1.5, 
we find that the flux at the Lyman limit should be of the order of 
$F_{\rm
o}$~=~5$\times$10$^{-20}$
or 1.5$\times$10$^{-19}$~erg~cm$^{-2}$~s$^{-1}$~Hz$^{-1}$~sr$^{-1}$
respectively.
This is much higher than the UV background from QSOs and early galaxies
at this redshift (Miralda-Escud\'e \& Ostriker 1990). 
Some additional source must
contribute the ionization. If this is the quasar, we derive that the distance
to the quasar should be larger than 
0.3$h^{-1}$~Mpc ($q_{\rm o}$~=~0.5), using a typical luminosity of 
$F_{\rm o}$~=~1.4$\times$10$^{-28}$~erg~cm$^{-2}$~s$^{-1}$~Hz$^{-1}$~sr$^{-1}$
(Sargent \& Steidel 1991). This strongly suggests that the system is not 
associated with the QSO. 
\section{Conclusion} \label{s5}
We have shown in the previous sections that the gas
lying on the line of sight to Tol~1037-2704 is certainly not associated 
with the 
QSO since the physical state is notably different from that of
BAL or associated systems (e.g. Wampler et al.
1995, Petitjean et al. 1994). \par
Most of the systems on the line of sight
have column densities and ionization states typical of 
intervening systems with, when an estimate is possible, abundances of 
the order of 10$^{-2}$--10$^{-1}$~$Z_{\odot}$. A detailed 
study of the O~{\sc i}, 
C~{\sc ii}$^*$, Si~{\sc ii} and Fe~{\sc ii} lines at
$z_{\rm abs}$~$\sim$~2.1393 shows that the oxygen 
metallicity is of the order of 0.03~$Z_{\odot}$ and
[Fe/O] is slightly larger than solar; the ionization parameter 
and electronic density are of the order of $U$~$\sim$~10$^{-3}$ and
$n_{\rm e}$~$\sim$~3~cm$^{-3}$. The distance to the QSO is probably 
larger than 300$h^{-1}$~kpc. The system at $z_{\rm abs}$~$\sim$~2.08
has a shallow and complex profile both on the line of sight to 
Tol~1037-2704 and Tol~1038-2712. 
The fact that the strengths of the two lines of the C~{\sc iv} and 
Si~{\sc iv} doublets are not in the oscillator strength ratios indicates
that the complex is made up of narrow unresolved components.
The temperature must be smaller than 10$^5$~K. The abundances
are larger than 0.1~$Z_{\odot}$. If the gas is photo-ionized, the
ionizing spectrum must be stellar. Alternatively, the gas
could be purely collisionally ionized. In this case however, silicon must
be overabundant relative to carbon by a factor of ten.\par
These results strengthen the conclusion
by Sargent \& Steidel (1987) and DI96 that the material is part of
an intervening supercluster. The latter authors have shown
that the number of C~{\sc iv} systems toward Tol~1037-2704 
and Tol~1038-2712 is
much in excess of what is expected. As we confirm seven of their probable 
systems toward Tol~1037-2704, the number of systems with
$w_{\rm r}$~$>$~0.15~\AA~ is equal to 16 to be compared with 2 expected
from Poisson statistics.
Considering coincidences of absorption redshifts in four lines of sight
in the same field, they find also that there is a marginally 
significant correlation on comoving scales of $<$~18~Mpc.\par
This altogether strongly suggests that we observe a supercluster
which extends over 30$\times$80$h^{-1}$~Mpc$^{2}$. 
Deep imaging should be performed in this field to detect the 
luminous component of this large scale structure.
\begin{acknowledgements} 
We would like to thank the referee, Chris Impey,
who carefully reviewed our manuscript.
\end{acknowledgements}

\end{document}